\documentclass[conference]{IEEEtran}
\usepackage{balance}

\usepackage{tikz}
\usetikzlibrary{arrows.meta,positioning,fit,shapes,backgrounds,calc,decorations.pathreplacing}
\usepackage{listings}
\usepackage{lscape}
\usepackage{minted}
\usepackage{dirtree}

\usepackage[inkscapeformat=png]{svg}

\usepackage{enumitem}
\usepackage{graphicx}
\usepackage{multicol, latexsym}
\newcommand{\clip}{\mathrm{clip}}
\usepackage{blindtext}
\usepackage[skip=5pt,font=small]{caption}
\usepackage{longtable}
\usepackage{csquotes}
\usepackage{amsfonts}
\usepackage{amsmath}
\usepackage{amsthm}
\usepackage{amssymb}
\usepackage{algorithm}
\usepackage{tabularx}
\usepackage{capt-of,lipsum} % dedicated caption for equation
\usepackage{subcaption}

\usepackage{listings}
\usepackage{caption}
\usepackage{comment}
\usepackage{diagbox}
\usepackage{pdflscape}
\usepackage{booktabs}
\usepackage{makecell}
\usepackage{multirow}
\usepackage{fourier} 
\usepackage{array}

\usepackage{siunitx}
\usepackage{mathtools, nccmath}

\usepackage{siunitx}
\DeclareSIUnit[number-unit-product = { }] \dBm{dBm}

\usepackage{algpseudocode}
\usepackage{algorithm}

\usepackage{listing-styles}
\usepackage{wasysym}

\algnewcommand\algorithmicforeach{\textbf{for each}}
\algdef{S}[FOR]{ForEach}[1]{\algorithmicforeach\ #1\ \algorithmicdo}

\usepackage[hidelinks]{hyperref}

\definecolor{myblue}{rgb}{0.09,0.20,0.34}
\definecolor{mygreen}{rgb}{0,0.6,0}
\definecolor{mygray}{rgb}{0.98,0.98,0.98}
\definecolor{myorange}{rgb}{0.92,0.49,0.34}
\definecolor{mywhite}{rgb}{1.0,1.0,1.0}

\definecolor{NMR}{RGB}{255,255,86}
\definecolor{MRA}{RGB}{255,231,27}
\definecolor{MRD}{RGB}{178,178,178}
\definecolor{MRIP}{RGB}{188,172,0}
\definecolor{MRS}{RGB}{161,207,106}
\definecolor{NA}{RGB}{228,60,52}
\definecolor{AI}{RGB}{255,255,86}
\definecolor{RMR}{RGB}{162,4,21}
\definecolor{RMD}{RGB}{178,178,178}
\definecolor{RMA}{RGB}{255,231,27}
\definecolor{RMIP}{RGB}{188,172,0}
\definecolor{RMS}{RGB}{161,207,106}
\definecolor{LIGHT_GREY}{RGB}{240,240,240}
\definecolor{lightgray}{rgb}{.9,.9,.9}
\definecolor{darkgray}{rgb}{.4,.4,.4}
\definecolor{purple}{rgb}{0.65, 0.12, 0.82}
\definecolor{darkgreen}{RGB}{30, 142, 20}

\newcommand{\etal}{\textit{et al}.\ }

\newcommand{\eg}{\textit{e}.\textit{g}.,\ }
\newcommand{\cf}{\textit{c}\textit{f}.\ }

\newsavebox{\mybox}

\usepackage{pifont}% http://ctan.org/pkg/pifont

\usepackage{censor}
\usepackage[scaled]{helvet}
\usepackage[T1]{fontenc}
\usepackage{helvet}

\usepackage{tabularx,booktabs,makecell,array}
\newcolumntype{Y}{>{\centering\arraybackslash}p{0.06\textwidth}} % compact centered yes/no cells

\begin{document}

% Paper Title: Titles are generally capitalized except for words such as a, an, and, as, at, but, by, for, in, nor, of, on, or, the, to and up, which are usually not capitalized unless they are the first or last word of the title.



%\title{Algorithmic Framework for Sound Localization in Polyphonic Noisy Environments}

%\title{Building a Distributed Pulse-Wave Simulator\\for DDoS Dataset Generation}
\title{CareNet: Linking Home-router Network Traffic to DSM-5 Depressive Behavior Indicators}

% author names and affiliations: use a multiple column layout for up to three different affiliations

\DeclareRobustCommand*{\IEEEauthorrefmark}[1]{%
  \raisebox{0pt}[0pt][0pt]{\textsuperscript{\footnotesize #1}}%
}

\author{\IEEEauthorblockN{ 
Stephan Nef,
Bruno Rodrigues
}

%BR: shall we use some kind of short form for affiliations?

%\IEEEauthorblockA{\IEEEauthorrefmark{1}Communication Systems Group CSG, Department of Informatics IfI, 
%University of Zurich UZH, Switzerland\\
%E-mail: [mueller|schumm|niu|stiller]@ifi.uzh.ch, stefanrichard.saxer@uzh.ch, bruno.rodrigues@unisg.ch, fraenan@upv.es
%}

\IEEEauthorblockA{Embedded Sensing Group ESG, School of Computer Science SCS, %Embedded Sensing Group ESG, Institute of Computer Science in Vorarlberg ICV,\\
University of St. Gallen HSG, Switzerland\\
%E-mail: bruno.rodrigues@unisg.ch
} 
E-mail: stephan.nef@student.unisg.ch|bruno.rodrigues@unisg.ch
}

% Define definition environment

\newtheoremstyle{mydef}
{\topsep}{\topsep}%
{}{}%
{\bfseries}{}
{\newline}
{%
  \rule{\linewidth}{0.4pt}\\*%
  \thmname{#1}~\thmnumber{#2}\thmnote{\ -\ #3}.\\*[-1.5ex]%
  \rule{\linewidth}{0.4pt}}%
\theoremstyle{mydef}
\newtheorem{definition}{Definition}
\newtheorem{protocol}{Step}

% create black circles with numbers in it
%\newcommand*\circled[1]{\tikz[baseline=(char.base)]{
%            \node[shape=circle,fill,inner sep=2pt] (char) {\textcolor{white}{#1}};}}
            
%\DeclareRobustCommand{\circled}[1]{\tikz[baseline=(myanchor.base)] \node[font=\tiny,circle,fill=.,inner sep=1pt] (myanchor) {\color{-.}\bfseries\footnotesize #1};}


\maketitle
% As a general rule, do not put math, special symbols or citations in the abstract

%%%%%%%%%%%%%%%%%%%%%%%%%%%%%%%%
% Recipe to do an abstract:
%%%%%%%%%%%%%%%%%%%%%%%%%%%%%%%%
%(1) Context
%(2) Research Gap
%(3) Goals, Objectives
%(4) Methodology, how
%(5) Results

% Example:
%Abstract
%(1) In the context of energy efficiency in computer networks, a significant number of solutions ranging from protocols and functionalities to energy efficiency-oriented management applications have been proposed. 
%(2) However, the characteristics of environments to develop and validate such solutions are not as discussed as the solutions themselves. 
%(3) Considering this, this work proposes an emulation environment to develop and validate energy efficiency-oriented solutions, as well as discuss their specific characteristics. 
%(4) Thus, three functionalities of different network scopes are implemented, Adaptive Link Rate (interface level), Syncronized Coalescing (device level) and SustNMS (network level) in the Mininet emulation environment using the implemented software-defined networks paradigm on the POX controller. 
%(5) The environment is validated by comparing the energy savings achieved by these features in a topology inspired by the National Research Network (RNP).

\begin{abstract}
Digital mental-health sensing increasingly depends on mobile or wearable devices that require intrusive permissions and continuous user compliance. We present \emph{CareNet}, a router-centric system that transforms household network metadata into interpretable behavioral indicators aligned with DSM-5 depressive-symptom domains. All processing occurs locally at the home gateway, preserving privacy while maintaining visibility of temporal routines.

The core contribution is the \emph{Fuzzy Additive Symptom Likelihood (FASL)}, a transparent formulation that fuses header-level metrics into daily criterion-level likelihoods using bounded fuzzy memberships and additive aggregation. Combined with a DSM-style temporal gate, FASL integrates short-term traffic fluctuations into persistent, clinically interpretable indicators. Evaluation on realistic multi-day traces shows that CareNet captures characteristic patterns such as delayed sleep timing and attentional instability without payload inspection. The results highlight the feasibility of reproducible, explainable behavioral inference from router-side telemetry.
\end{abstract}

\begin{IEEEkeywords}
Router-centric sensing, network metadata, DSM-5, Behavior Observation Metrics
\end{IEEEkeywords}

% For peer review papers, you can put extra information on the cover page as needed:

% \fi
%
% For peer review papers, this IEEEtran command inserts a page break and creates the second title. It will be ignored for other modes.

\IEEEpeerreviewmaketitle

\pagestyle{plain} % adds with ^ numbering to pages

% trigger a \newpage just before the given reference number - used to balance the columns on the last page adjust value as needed - may need to be readjusted if the document is modified later
% The "triggered" command can be changed if desired: \IEEEtriggercmd{\enlargethispage{-5in}}
%====================================================
% Paper sections to be included in this very order 

\section{Introduction}
Depression is a leading contributor to the global burden of disease and often begins in adolescence, where it predicts lower educational attainment and poorer adult outcomes \cite{mcgorry2018early,clayborne2019systematic}. Early intervention matters, yet delays from symptom onset to evidence‑based care are common and worsen prognosis \cite{oguchi2014delay,mcgorry2018early}. Digital behavioral monitoring has linked passively collected signals to sleep, activity, and social rhythm, but device‑centric approaches require apps, ongoing compliance, and invasive permissions; in multi‑person households they are unreliable and raise privacy and deployment concerns \cite{saeb2015mobile,wang2014studentlife,ware2020predicting}.

\begin{figure}[t]
    \centering
    \includegraphics[width=1\linewidth]{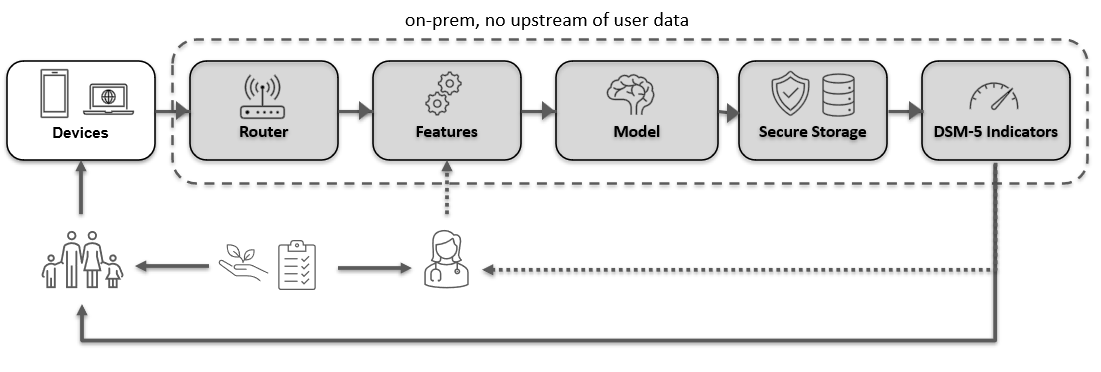}
    \caption{Overview of CareNet's approach: From parsing network traffic and extracting features to mapping features into BOMs and DSM-aligned indicators. CareNet's source-code is available: \cite{nef_app_pcap_to_dsm5_public}}
    \label{fig:carenet-overview}
\end{figure}

This paper investigates to which extent a router-centric vantage point can provide meaningful insights on mental-health awareness. The central idea (\cf in Figure~\ref{fig:carenet-overview}) is to analyze header-level timing and coarse destination information (\eg Public Suffix List and IAB - Interactive Advertising Bureau - content taxonomy labels) without inspecting payloads, and to summarize these signals into \emph{Behavior Observation Metrics (BOMs)} that align with symptom patterns described in DSM-5 (Diagnostic and Statistical Manual of Mental Disorders) \eg disturbed sleep–wake rhythm, social withdrawal \cite{publicsuffix,iabtaxonomy,americanpsychiatricassociation2022dsm5}. All processing remains local to avoid data exposure and operational overhead.

\begin{table*}[ht]
\centering
\footnotesize
\setlength{\tabcolsep}{2pt}
\renewcommand{\arraystretch}{1.08}
\begin{tabularx}{\textwidth}{@{}>{\raggedright\arraybackslash}p{0.21\textwidth} p{0.35\textwidth} *{6}{Y}@{}}
\toprule
\textbf{Related} & \textbf{Short description} &
\textbf{Router} & \textbf{Meta} & \textbf{Attr.} & \textbf{Interp.} & \textbf{DSM gate} & \textbf{Privacy} \\
\midrule
Wang \etal (2014) \cite{wang2014studentlife} & Smartphone app (sleep, mobility, social rhythm) & \Circle & \Circle & \LEFTcircle & \LEFTcircle & \Circle & \Circle \\
Saeb \etal (2015) \cite{saeb2015mobile} & Phone GPS/accelerometer for PHQ-9 correlation & \Circle & \Circle & \LEFTcircle & \Circle & \Circle & \Circle \\
Wang \etal (2018) \cite{wang2018tracking} & Phone + Fitbit multimodal fusion & \Circle & \LEFTcircle & \LEFTcircle & \LEFTcircle & \Circle & \LEFTcircle \\
Saeb \etal (2017) \cite{saeb2017passive} & Mobile passive sleep monitoring & \Circle & \Circle & \LEFTcircle & \Circle & \Circle & \Circle \\
Kushlev \etal (2016) \cite{kushlev2016silence} & Notification bursts and attention effects & \Circle & \LEFTcircle & \LEFTcircle & \LEFTcircle & \Circle & \LEFTcircle \\
Alamoudi \etal (2024) \cite{alamoudi2024evaluating} & Sleep-tracker app, anxiety/depression prediction & \Circle & \Circle & \LEFTcircle & \LEFTcircle & \Circle & \Circle \\
Cacheda \etal (2019) \cite{cacheda2019early} & Social-media text mining (Twitter) & \Circle & \Circle & \LEFTcircle & \Circle & \Circle & \Circle \\
De Choudhury \etal (2016) \cite{dechoudhury2016shift} & Reddit linguistic shift analysis & \Circle & \Circle & \LEFTcircle & \LEFTcircle & \Circle & \Circle \\
Bucur \etal (2024) \cite{bucur2024leveraging} & LLM-based social-media depression detection & \Circle & \Circle & \LEFTcircle & \Circle & \Circle & \Circle \\
Ware \etal (2018) \cite{ware2018large} & Campus Wi-Fi metadata for screening & \LEFTcircle & \CIRCLE & \LEFTcircle & \LEFTcircle & \Circle & \LEFTcircle \\
Mammen \etal (2021) \cite{mammen2021wisleep} & Home-Wi-Fi sleep duration inference & \CIRCLE & \CIRCLE & \LEFTcircle & \LEFTcircle & \Circle & \CIRCLE \\
Ware \etal (2020) \cite{ware2020predicting} & Smartphone network usage predicting mood & \CIRCLE & \CIRCLE & \Circle & \LEFTcircle & \Circle & \CIRCLE \\
Stachl \etal (2020) \cite{stachl2020predicting} & Large-scale smartphone metadata, personality traits & \Circle & \LEFTcircle & \LEFTcircle & \LEFTcircle & \Circle & \LEFTcircle \\
\textbf{CareNet (this paper)} & Router-centric metadata for behavioral indicators & \CIRCLE & \CIRCLE & \CIRCLE & \CIRCLE & \CIRCLE & \CIRCLE \\
\bottomrule
\end{tabularx}
\caption{Representative related work and their alignment with categories that are central. Legend: \CIRCLE\,= provided; \LEFTcircle\,= partially;
\Circle\,= not provided.}
\label{tab:rw-compare-full}
\end{table*}

We consider a multi-person household in which all devices share a single home gateway. In this setting, we propose \textit{CareNet}, an end-to-end pipeline (\cf Figure~\ref{fig:carenet-overview}) that: (i) captures minute level network activity and enriches flows with coarse domain labels; (ii) derives interpretable features such as first online time, nocturnal bursts, protocol diversity, and repetition from packet headers only; (iii) groups related features into \emph{Behavior Observation Metrics} (BOMs) with explicit \emph{risk direction} and clinical rationale; (iv) fuses these into \emph{criterion level likelihoods} through transparent fuzzy memberships; and (v) summarizes rolling histories using a DSM-style gate to produce day level indicators. The design favors transparency and auditability over black-box accuracy and confines observation to non content metadata for privacy preserving and interpretable insight.

Four main challenges arise in this scenario. The first is \textit{temporal dilution}, where short but behaviorally significant deviations (\eg an unusually late night of activity or an evening of social withdrawal) vanish when data are averaged into daily or weekly aggregates \cite{saeb2015mobile,wang2018tracking,hale2015screentime}. The second is \textit{multi-user mixtures}, since household traffic merges multiple devices and users, obscuring individual changes \cite{ware2018large,mammen2021wisleep}. The third concern \textit{underused header dynamics}: router-level studies often reduce traffic to byte counts, overlooking informative timing, burst, and destination patterns that could reveal behavior without content inspection \cite{ware2018large}. Finally, the fourth challenge is the \textit{adoption barrier} of opaque models, as lack of interpretability hinders clinical trust and calls for transparent mappings from features to DSM-aligned indicators \cite{svenaeus2014diagnosing,pedersen2001icd}.

CareNet makes the following contributions:
\begin{itemize}
\item \textbf{Router-centric telemetry:} a household-gateway vantage that extracts behavioral signals from header metadata while preserving privacy.
\item \textbf{Behavior Observation Metrics (BOMs):} a criterion-aligned mapping from network timing and domain diversity to interpretable behavioral indicators.
\item \textbf{Fuzzy Additive Symptom Likelihood (FASL):} a transparent fusion model combining fuzzy memberships and a DSM-style gate to derive daily DSM-5–aligned likelihoods.
\item \textbf{Prototype and evaluation:} an end-to-end implementation on real traces demonstrating feasibility, stability, and reproducibility under privacy constraints.
\end{itemize}

Our target setting is adolescents in multi‑person households. Indicators are \emph{non‑diagnostic} and intended to support early observation by caregivers and clinicians; payloads are not inspected, and processing can remain local. The work complements (not replaces) established tools such as PHQ‑9 \cite{kroenke2001phq,smith2018validation}.

The paper is organized as follows. Section~\ref{sec:background} reviews background and related work. Section~\ref{sctn::design} details the \textit{CareNet} design and FASL formulation. Section~\ref{sec:evaluation} reports the evaluation, and Section~\ref{sec:final_considerations} concludes the paper.

%Section~\ref{sec:background} summarizes clinical and networking background and privacy constraints. Section~\ref{sec:related} positions the work against device‑centric phenotyping and router‑level inference. Section~\ref{sec:method} details BOM design, fuzzy memberships, and the DSM‑style gate. Section~\ref{sec:implementation} describes the router‑centric architecture and dataflow. Section~\ref{sec:evaluation} evaluates feasibility on realistic traces. Section~\ref{sec:discussion} discusses limitations and future work; Section~\ref{sec:conclusion} concludes.

%=====================================================================
\section{Background: DSM-5 and BOM} \label{sec:background}
%=====================================================================

\noindent \textbf{DSM-5} defines a \emph{Major Depressive Disorder (MDD)} as at least five of nine core symptoms persisting for a minimum of two weeks, with one being either depressed mood or loss of interest~\cite{svenaeus2014diagnosing,americanpsychiatricassociation2022dsm5}. \textit{CareNet} follows this definition to align all behavioral indicators with clinically recognized criteria while preserving privacy and interpretability (Table~\ref{tab:dsm5-symptoms}).

\begin{table}[h]
\centering
\caption{DSM-5 symptom indicators for MDD \cite{americanpsychiatricassociation2022dsm5}.}
\label{tab:dsm5-symptoms}
    \begin{tabular}{cl}
    \toprule
    \textbf{ID} & \textbf{Symptom Indicator} \\
    \midrule
    (1) & Depressed mood \\
    (2) & Loss of interest or pleasure (anhedonia) \\
    (3) & Significant weight change or appetite disturbance \\
    (4) & Insomnia or hypersomnia \\
    (5) & Psychomotor retardation or agitation \\
    (6) & Fatigue or loss of energy \\
    (7) & Feelings of worthlessness or excessive guilt \\
    (8) & Diminished ability to think, concentrate, or indecisiveness \\
    (9) & Recurrent thoughts of death or suicidality \\
    \bottomrule
    \end{tabular}
\end{table}

\noindent \textbf{Behavior Observation Metric (BOM)} acts as a digital behavioral biomarker that aggregates observable network and environmental features into interpretable indicators aligned with DSM-5 symptom domains~\cite{saeb2015mobile,wang2014studentlife}. While not diagnostic, it translates feature-level evidence into criterion-level likelihoods for early awareness. Each BOM emphasizes \textit{validity} (explicit mapping to DSM-5 constructs), \textit{reliability} (temporal stability and sensitivity to change), and \textit{transparency} (auditable memberships, weights, and gate parameters within the framework).

\section{Related work} \label{sec:related-work}

Table~\ref{tab:rw-compare-full} provides a summary of each related work, highlighting the focus or modality. %The columns capture, in turn: a router‑centric vantage (signals observed at the home gateway rather than only on devices); metadata‑only sensing (header‑level timing and coarse destinations without payload inspection); person‑level attribution in multi‑user homes; transparent or otherwise inspectable modeling; an explicit DSM‑style decision gate that aggregates evidence at the criterion level; and privacy‑by‑design choices such as local processing and minimal retention.
Device-centric studies using apps or wearables show digital traces correlate with sleep, mobility, and mood \cite{wang2014studentlife,saeb2015mobile,wang2018tracking,saeb2017passive}. However, they score low on router-centric and metadata-only axes due to reliance on per-device instrumentation and permissions. Person attribution assumes one-device–one-user, complicating issues like short behavioral excursions being diluted by long windows and data fragility with low compliance. Models are either partly interpretable or opaque, and lack a formal DSM-style aggregation step.

A second line of work analyzes social‑media content to detect depressive language \cite{cacheda2019early, dechoudhury2016shift, bucur2024leveraging}. These approaches achieve strong performance in text domains, but by design they violate the metadata‑only constraint and carry privacy trade‑offs; interpretability varies and criterion‑level decision logic is typically absent. As a result, they are not suitable for a home‑gateway that must avoid payload inspection.

Network-centric environmental sensing aligns with our goals, as WiFi studies \cite{ware2018large, mueller2024zigbee,mammen2021wisleep, ware2020predicting} achieve a full circle in metadata-only sensing and router vantage progress, revealing sleep and mobility patterns via network signatures. These studies, however, typically focus on aggregate households or single behaviors, leaving person-level attribution unresolved, interpretability partial, and lacking an explicit DSM-style gate. Additionally, the underutilization of rich header dynamics like burst structure, protocol mix, and repetition limits the clinical feature mapping.

\emph{CareNet} tackles challenges in various categories such as temporal dilution and multi-user mixtures by using a router-centric, metadata-only perspective. It introduces person-level attribution at gateways, enhances transparency through interpretable \emph{Behavior Observation Metrics}, and uses a DSM-style gate for daily likelihood summaries while maintaining privacy. This approach complements existing device-centric and content-based methods.

\section{Methodology}
\label{sctn::design}

\begin{figure}[h]
    \centering
    \includegraphics[width=1\linewidth]{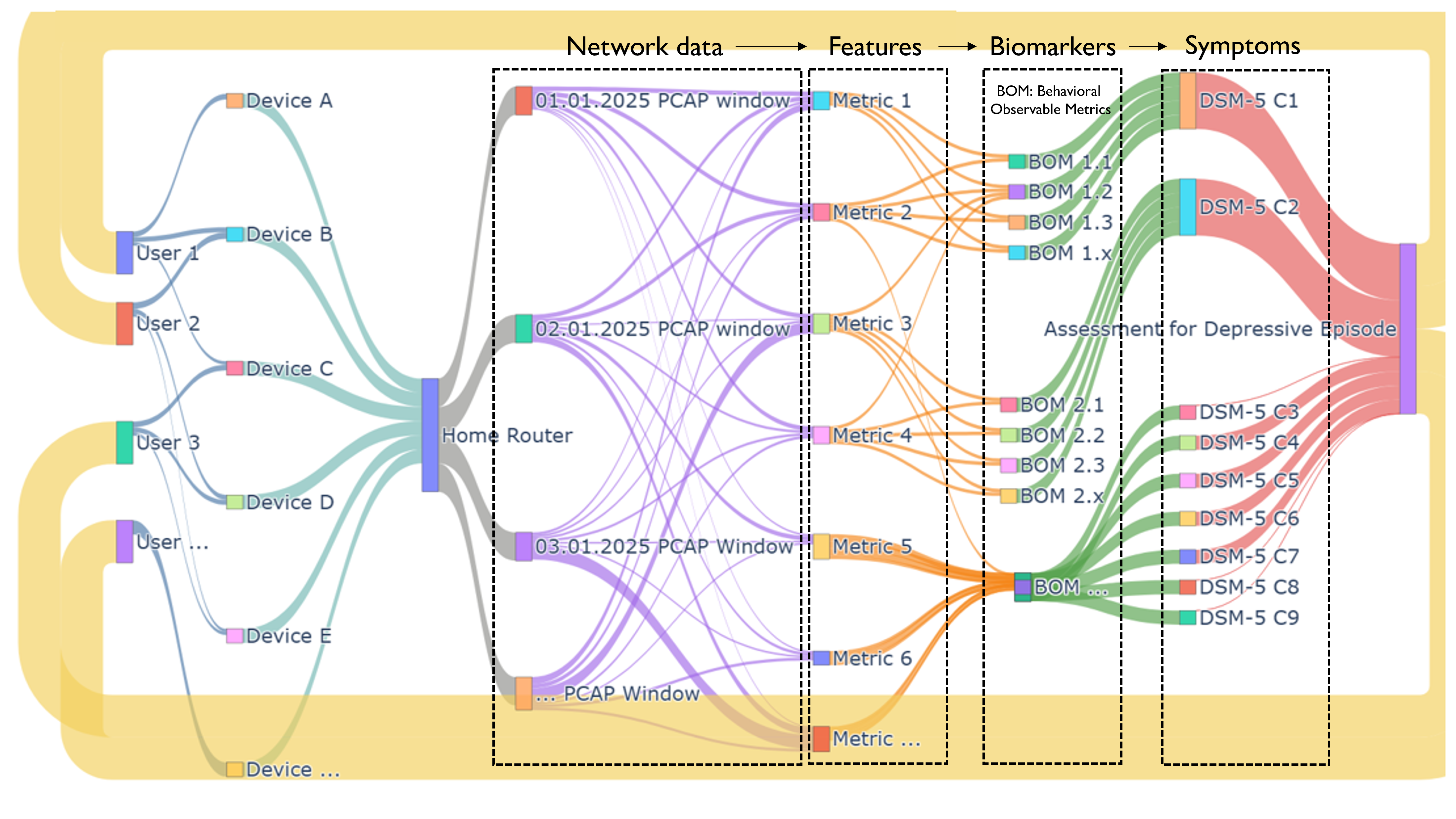}
    \caption{Sankey diagram illustrating CareNet's general workflow}
    \label{fig:carenet-sankey}
\end{figure}

\textit{CareNet} is designed as a router-centric, privacy-first system that interprets network metadata within the household to infer behavioral metrics relevant to mental-health awareness. Figure \ref{fig:carenet-sankey} summarizes our approach and the closed information‑flow loop that makes router‑centric observation useful over time.

 %Main components are described as follows:
%\begin{itemize}
%    \item \textbf{Gateway/Router}: central point capturing household network metadata.  
%    \item \textbf{Network capture}: collects packet headers and timing without payloads.  
%    \item \textbf{Feature engineering}: derives daily behavioral features from metadata.  
%    \item \textbf{Behavior Observation Metric (BOM)}: groups features into interpretable behaviors.  
%    \item \textbf{DSM-5 mapping}: aligns behaviors with DSM-5 symptom criteria.  
%    \item \textbf{DSM-5 gating and likelihood}: aggregates daily evidence into criterion presence.  
%\end{itemize}

\begin{figure}[t]
  \centering
  % Draws a frame around the entire figure
  %\fcolorbox{black}{white}{%
    %\begin{minipage}{0.93\columnwidth}
      \centering
      % Resize to fit inside the frame
      \resizebox{!}{1.2\columnwidth}{%
        \begin{tikzpicture}[
          node distance=7mm and 16mm,
          every node/.style={font=\small},
          box/.style={draw, rounded corners, align=center, inner sep=4pt,
                      minimum width=40mm, minimum height=9mm, fill=gray!5},
          store/.style={draw, cylinder, shape border rotate=90,
                        aspect=0.25, minimum height=12mm, minimum width=28mm,
                        align=center, fill=gray!5},
          note/.style={draw, rounded corners, align=left, inner sep=4pt,
                       minimum width=46mm, minimum height=10mm, fill=green!7},
          arrow/.style={-{Latex[length=2.5mm]}, thick}
        ]

          % ====== MAIN VERTICAL PIPELINE ======
          \node[box] (router) {Home Router};
          \node[box, right=of router] (gw) {Gateway};
          \node[box, below=of gw] (pcap) {Network Capture\\(PCAP / tcpdump)};
          \node[box, below=of pcap] (parser) {ETL Parser \& Filtering\\time, IPs, ports, protocol, bytes, DNS/SNI};
          \node[store, below=of parser] (parts) {Partitioned Storage\\$n$-minute segments};
          \node[box, below=of parts] (features) {Feature Extraction\\digital traffic + environmental signals};
          \node[store, below=of features] (fstore) {Feature Store\\per-segment records};
          \node[box, below=of fstore] (feng) {Feature Engineering};
          \node[box, below=of feng] (bom) {Behavior Observation Metric};
          \node[box, below=of bom] (crit) {DSM-5 Criterion Modules\\C1–C9 daily likelihoods $L_k$};
          \node[box, below=of crit] (gate) {DSM Gate\\window $M$, days $N$, threshold $\theta$};
          \node[box, below=of gate] (episode) {Depression Episode Likely};
          \node[note, below=of episode] (ui) {Interactive Indicator Dashboard};

          % Main arrows
          \draw[arrow] (router) -- (gw);
          \draw[arrow] (gw) -- (pcap);
          \draw[arrow] (pcap) -- (parser);
          \draw[arrow] (parser) -- (parts);
          \draw[arrow] (parts) -- (features);
          \draw[arrow] (features) -- (fstore);
          \draw[arrow] (fstore) -- (feng);
          \draw[arrow] (feng) -- (bom);
          \draw[arrow] (bom) -- (crit);
          \draw[arrow] (crit) -- (gate);
          \draw[arrow] (gate) -- (episode);
          \draw[arrow] (episode) -- (ui);

          % ====== USER PROFILES & IDENTITY RESOLUTION (RIGHT SIDE) ======
          \node[box, right=25mm of pcap] (discover) {Device \& Session Discovery\\addresses, identifiers, DNS/SNI};
          \node[box, below=of discover] (idmap) {Identity Resolution\\IP to Profile allocation};
          \node[store, below=of idmap] (profiles) {User Profile Registry};

          % Connections
          \draw[arrow] (discover) -- (idmap);
          \draw[arrow] (idmap) -- (profiles);
          \draw[arrow] (profiles) -- (fstore);

          % ====== PARAMETERS / CACHE (RIGHT SIDE) ======
          \node[store, right=25mm of feng] (params) {Parameter Registry};
          \node[store, right=25mm of fstore] (cache) {Feature Cache};

          \draw[arrow] (params.west) -- (feng.east);
          \draw[arrow] (params.south west) to[bend right=20] (crit.east);
          \draw[arrow] (fstore.east) -- (cache.west);
          \draw[arrow] (cache.south west) to[bend left=18] (feng.east);

          % ====== PRIVACY & PERSONALIZATION REGIONS ======
          \node[draw, dashed, rounded corners,
                fit=(pcap)(parser)(parts)(features)(fstore),
                label={[align=right]left:Privacy-first\\local processing\\data minimization\\retention policy\\aggregate outputs}] {};

          \node[draw, dashed, rounded corners,
                fit=(parser)(profiles)(feng)(episode),
                label={[align=right]left:Personalization\\per-user weights\\profile-aware thresholds}] {};

        \end{tikzpicture}
      }
    %\end{minipage}
  %}

  \caption[System Design.]{\textit{CareNet} detailed workflow, detailing key components and the flow through parsing, storage, feature extraction, and BOM aggregation toward DSM-5 MDD indicators and an interpretable gate.}
  \label{fig:carenet-design}
\end{figure}
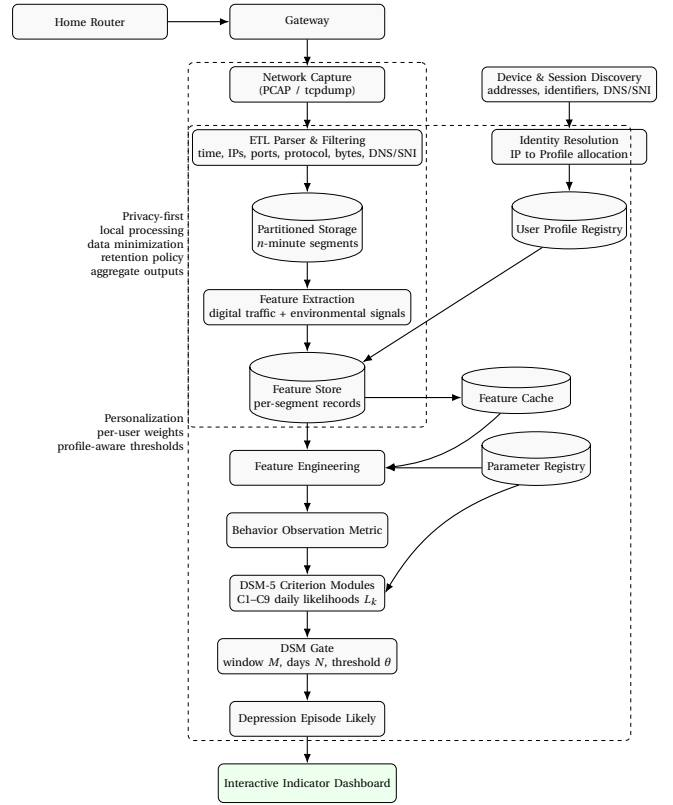

Figure~\ref{fig:carenet-design} illustrates the end-to-end workflow. Traffic from all devices passes through the \textit{Home Router} to the \textit{Gateway}, where \textit{Network Capture} records header metadata, the \textit{ETL Parser} (Extract, Transform, Load) extracts key fields, and \textit{Partitioned Storage} organizes them into short, fixed segments for local, privacy-preserving analysis.

%Traffic from all connected devices on the home network first reaches \textbf{Home Router} and is mirrored at \textbf{Gateway}, where \textbf{Network Capture (PCAP / tcpdump)} records packet information. An \textbf{ETL Parser \& Filtering} stage extracts timestamps, addresses, ports, protocol bytes, and, when available, DNS or SNI hostnames (not limited to it). Under the privacy-first regime indicated by the dashed boundary, records are organized in  \textbf{Partitioned Storage} as fixed segments of $n$-minutes that allow local processing, data minimization, and short retention before any analysis proceeds.

In parallel (\cf Figure \ref{fig:carenet-design} right upper side), the system maintains an inventory of devices via \textit{Device \& Session Discovery} and resolves household ambiguity through \textit{Identity Resolution (IP to Profile allocation)}, storing assignments in \textit{User Profile Registry}. These identity services feed the analytical path at two points. First, they inform \textit{Feature Extraction} so that the measurements per-segment reflect the intended individual, and second, condition \textit{Feature Store}, where the records per-segment are retained. A lightweight \textit{Feature Cache} accelerates iterative analysis and a \textit{Parameter Registry} maintains analyst-controlled settings for memberships, weights, and gate parameters.

From the feature store, the measures aligned with the criterion are calculated in \textit{Feature Engineering} and grouped into interpretable components within \texttt{BOM}. The \textit{DSM-5 MDD Criterion Modules (C1-C9)} transform these components into daily probables based on a possible criterion $L_k \in [0,1]$. The \textit{DSM Gate} then applies a clinically motivated rule on a rolling window $M$: a criterion is marked present when $L_k \ge \theta$ on at least $N$ days. %However, the design that is not restricted to this mechanism means that any other transparent and clinically interpretable model that fulfills the requirements could be integrated at this stage.

\subsection{Mapping Network Features to Behavioral Metrics}

Table~\ref{tab:network_feature_mapping} summarizes how (selected) router-visible network features are grouped into interpretable \emph{Behavior Observation Metrics (BOM)} and aligned with DSM-5 symptom domains. Each DSM‑5 criterion (C1–C9) (\cf Table \ref{tab:dsm5-symptoms} on DSM-5 MDD symptoms) is bolstered by a concise array of criteria-aligned metrics that convert router-side measurements into interpretable signaling indicators with a clear orientation (\emph{pro} or \emph{contra}).

System‑design‑wise, metrics are intermediate objects in a structured mapping: related metrics are grouped into clinically meaningful BOM, and BOM are mapped to criterion‑level likelihoods. This separation (metrics $\rightarrow$ BOM $\rightarrow$ criterion) keeps clinical meaning visible, enables profile‑aware tuning through a parameter registry, and makes the pathway from quantitative network data to qualitative constructs auditable. 

\begin{table}[t]
  \centering
  \caption{Feature mapping specification with descriptor/type.}
  \label{tab:network_feature_mapping}
  \scriptsize
  \setlength{\tabcolsep}{3pt}
  \renewcommand{\arraystretch}{1.05}
  \resizebox{\columnwidth}{!}{%
  \begin{tabular}{l l l l l c l l}
    \toprule
    \textbf{Network feature (selected examples)} & \textbf{Biomarker (BOM)} & \textbf{Ind.} & \textbf{Rel.} & \textbf{Dir.} & \textbf{W} & \textbf{Desc.} & \textbf{Signal group} \\
    \midrule
    % ===== C1: Depressed Mood =====
    Distinct social domains (eTLD+1) & Reduced social interaction & (1) & Gradual & --  & 1.0 & HLD & Service \\
    Median chat session duration & Reduced social interaction & (1) & Gradual & --  & 1.0 & LLD & Session \\
    Long streaming session duration & Passive media binge & (1) & Gradual & +   & 1.0 & LLD & Session \\
    Down/Up byte ratio ($D/U$) & Passive media binge & (1) & Gradual & +   & 1.0 & HLD & Volume \\
    Revisit diversity (Shannon index) & Rumination loops & (1) & Gradual & --  & 1.0 & HLD & Navigation \\
    Short-interval repeated visits & Rumination loops & (1) & Gradual & +   & 1.0 & LLD & Navigation \\
    Hourly traffic coefficient of variation & Flattened diurnal rhythm & (1) & Gradual & --  & 1.0 & HLD & Rhythm \\
    Night/Day traffic ratio & Flattened diurnal rhythm & (1) & Gradual & +   & 1.0 & HLD & Rhythm \\

    \midrule
    % ===== C2: Anhedonia =====
    Domain diversity (count eTLD+1) & Interest breadth & (2) & Gradual & --  & 1.0 & HLD & Service \\
    Category entropy (IAB topic mix) & Interest breadth & (2) & Gradual & --  & 1.0 & HLD & Service \\
    Chat session count (per day) & Engagement & (2) & Gradual & --  & 1.0 & HLD & Session \\
    Reply latency (chat / messaging) & Engagement & (2) & Gradual & +   & 1.0 & LLD & Interaction \\
    Passive/active ratio ($D/U$ vs.\ upload bursts) & Passivity & (2) & Gradual & +   & 1.0 & HLD & Volume \\
    Upload-burst rate per active hour & Passivity & (2) & Gradual & --  & 1.0 & HLD & Interaction \\

    \midrule
    % ===== C3: Appetite / Weight =====
    Delivery activity deviation (z-score) & Ordering activity & (3) & Gradual & both & 1.0 & HLD & Service \\
    Late-night delivery share ($>$23{:}00) & Ordering timing & (3) & Gradual & +   & 1.0 & HLD & Rhythm \\
    Diet/nutrition domain exposure & Diet focus & (3) & Gradual & both & 1.0 & HLD & Service \\

    \midrule
    % ===== C4: Sleep =====
    Digital sleep onset (min after 22{:}00) & Shifted sleep timing & (4) & Gradual & +   & 1.0 & HLD & Rhythm \\
    Onset-time variability (14-day circular SD) & Shifted timing irregularity & (4) & Gradual & +   & 1.0 & HLD & Rhythm \\
    Main nightly idle-gap length & Sleep duration change & (4) & Gradual & both & 1.0 & HLD & Rhythm \\
    Nocturnal micro-session count & Sleep fragmentation & (4) & Gradual & +   & 1.0 & LLD & Rhythm \\
    Inter-awakening gap (median) & Sleep fragmentation & (4) & Gradual & --  & 1.0 & LLD & Rhythm \\
    Daytime idle ratio (08--18\,h) & Daytime hypersomnia / flattening & (4) & Gradual & +   & 1.0 & HLD & Rhythm \\
    Night/Day traffic ratio & Flattened rhythm & (4) & Gradual & +   & 1.0 & HLD & Rhythm \\

    \midrule
    % ===== C5: Psychomotor =====
    Wi‑Fi re-associations/DHCP renewals (count) & Restlessness (device churn) & (5) & Gradual & +   & 1.0 & LLD & Mgmt \\
    Very short sessions ($<15$\,s) count & Restlessness (micro-activity) & (5) & Gradual & +   & 1.0 & LLD & Session \\
    Median inter-session gap & Motor slowing vs.\ agitation & (5) & Gradual & both & 1.0 & HLD & Session \\

    \midrule
    % ===== C6: Fatigue / Energy =====
    Midday idle minutes share & Low energy & (6) & Gradual & +   & 1.0 & HLD & Rhythm \\
    Daytime session count & Low energy & (6) & Gradual & --  & 1.0 & HLD & Session \\
    Sent bytes per active hour & Effortful interaction & (6) & Gradual & --  & 1.0 & HLD & Volume \\
    Upload-burst rate per active hour & Effortful interaction & (6) & Gradual & --  & 1.0 & HLD & Interaction \\
    Inter-click interval (mean/variance) & Slow browsing tempo & (6) & Gradual & +   & 1.0 & LLD & Interaction \\

    \midrule
    % ===== C7: Worthlessness / Guilt =====
    Mental-health resource domains (visits) & Help-seeking / self-worth & (7) & Gradual & +   & 1.0 & HLD & Service \\
    Therapist directory domains (visits) & Help-seeking & (7) & Gradual & +   & 1.0 & HLD & Service \\

    \midrule
    % ===== C8: Concentration / Decision =====
    Median page dwell time & Fragmented focus & (8) & Gradual & --  & 1.0 & LLD & Navigation \\
    DNS lookup burst rate (tab-hopping) & Fragmented focus & (8) & Gradual & +   & 1.0 & LLD & Navigation \\
    Notification-triggered micro-sessions & Interruptions & (8) & Gradual & +   & 1.0 & LLD & Session \\
    Back-navigation rate after click & Indecisive search & (8) & Gradual & +   & 1.0 & HLD & Navigation \\
    Search back-to-landing within 60\,s & Indecisive search & (8) & Gradual & +   & 1.0 & HLD & Navigation \\
    SERP time-to-first-click (domain change) & Indecision latency & (8) & Gradual & +   & 1.0 & LLD & Navigation \\

    \midrule
    % ===== C9: Suicidal Ideation =====
    Crisis-line domains (visits) & Crisis seeking & (9) & Gradual & +   & 1.0 & HLD & Service \\
    Self-harm community domains (visits) & Self-harm exposure & (9) & Gradual & +   & 1.0 & HLD & Service \\
    Cloud-backup surge (GB/day) & Digital affairs / planning & (9) & Gradual & +   & 1.0 & HLD & Volume \\
    \bottomrule
  \end{tabular}}
  \vspace{2pt}

  \raggedright\footnotesize
  \textit{Notes:} Ind.\ numbers = DSM-5 MDD criteria (1--9). Dir.\ $\in \{+,\;-,\;\text{both}\}$ denotes directionality (positive, negative, both).\\
  HLD = daily/high-level aggregate; LLD = low-level/session or burst measure. Service grouping can use eTLD+1 via PSL and optional IAB topic labels (header-only; content not inspected).\\[-2pt]
\end{table}

Network features are computed inside the privacy boundary from header/timing metadata only; registrable domains use eTLD+1 (Effective Top-Level Domain plus one), derived from the Public Suffix List (PSL) \cite{publicsuffix,iabtaxonomy}, and optional topical labels are header‑only (\cf Table \ref{tab:network_feature_mapping} notes). All metrics are daily aggregates or summaries produced from $n$‑minute partitions and written to the Feature Store, enabling local, profile‑aware processing via the Parameter Registry.

%For each behavioral component \(B_j(t)\), the system aggregates the standardized network features that belong to it using a weighted additive model (detailed in the next subsection). 
Each feature \(x_i\) is first normalized against the user's personal baseline to obtain a deviation \(z_{i,t}\), and then passed through a monotonic mapping function \(g_i(\cdot)\) 
that expresses the directionality of the feature (\emph{pro} or \emph{contra}) with respect to the behavioral construct. 

\subsection{Fuzzy Additive Symptom Likelihood (FASL)}

To convert packet-level network features into interpretable DSM-5 MDD indicators, we introduce the \emph{Fuzzy Additive Symptom Likelihood} (FASL). FASL treats traffic-derived metrics (\eg flow duration, idle-gap length, byte ratios) as noisy network measurements and fuses them into daily, criterion-level likelihoods. All this process is done at the edge (gateway) and realized in the steps as follows.

\noindent \textbf{(1) Triangular memberships.} Each network metric \(x_i(t)\) obtained from router telemetry or session statistics is mapped to \([0,1]\) using a simple triangular membership \(\mu_i(\cdot)\):
\begin{equation}
\mu_{\text{tri}}(x;\ell,m,h)=
\begin{cases}
0, & x \le \ell,\\
\dfrac{x-\ell}{m-\ell}, & \ell < x \le m,\\
\dfrac{h-x}{h-m}, & m < x < h,\\
0, & x \ge h.
\end{cases}
\label{eq:triangular_membership}
\end{equation}

%If $\ell=m$ or $m=h$, a one-sided shoulder
%($\mu_{\text{left}}$ or $\mu_{\text{right}}$) is used to avoid zero-width ramps;
%formal definitions appear in Appendix~\ref{app:fasl}.

The triangular membership is an ideal fit as it offers a computationally efficient and fully interpretable mapping with finite support in \([\ell,h]\), which is essential for continuous edge-level processing on the gateway. The function transforms a normalized network metric (\eg packet-rate deviation) into bounded evidence with finite support, making threshold logic explicit and verifiable.

\noindent  \textbf{(2) Biomarkers or Behavioral Observable Metrics.} Related network features form a signed component whose evidence remains bounded and interpretable:
\begin{equation}
S_{k,b}(t)
=\clip_{[-1,1]}\!\left(\sum_{i\in\mathcal{I}_b} w_{b,i}\, s_{k,i}\, \mu_{k,i}\!\big(x_i(t)\big)\right),
\quad \sum_i w_{b,i}=1.
\label{eq:signed-slice-compact}
\end{equation}

Features assigned to behavioral component \(b\) (for DSM-5 criterion \(k\)) are fused additively with normalized weights. Packet-level metrics from the same traffic dimension (\eg session timing or domain diversity) are fused using normalized weights, where \(s_{k,i}\in\{+1,-1\}\) encodes pro/contra orientation so that upload scarcity and burstiness, for example, can have opposite signs.

\noindent\textbf{(3) Criterion likelihood.} Signed component evidence $S_{k,b}\!\in[-1,1]$ is first mapped to a non-negative domain and then fused into a daily likelihood:
\begin{align}
\hat S_{k,b}(t) &= \tfrac{1}{2}\big(S_{k,b}(t)+1\big)\in[0,1], \\
L_k(t)
&=\clip_{[0,1]}\!\left(\sum_{b\in\mathcal{B}_k}
v_{k,b}\, \hat S_{k,b}(t)\right),
\qquad \sum_b v_{k,b}=1.
\label{eq:criterion-likelihood-compact}
\end{align}
The affine map $\tfrac{1}{2}(S+1)$ prevents silent negative-sum collapses and preserves interpretability as evidence mass. 
%Full derivation and clip definition are given in Appendix~\ref{app:fasl}.

\noindent \textbf{(4) DSM‑style gate.} A day is positive for criterion \(k\) if the likelihood exceeds a threshold; presence requires at least \(N\) positives in the last \(M\) days:
\begin{equation}
I_k(d)=\mathbb{I}[\,L_k(d)\ge\theta\,], 
\qquad
\mathrm{present}_k(t)=\mathbb{I}\!\left[\sum_{d=t-M+1}^{t} I_k(d)\ge N\right].
\label{eq:gate-compact}
\end{equation}

The gate applies a temporal sliding window over daily likelihoods, which is analogous to a moving-average filter in network monitoring—to require that signals persist over \(N\) of \(M\) days before being flagged.

\noindent  \textbf{(5) Episode rule.} Following DSM‑5, an episode is likely when \(\ge 5\) criteria are present, including \(\ge 1\) core (C1 or C2):
\begin{equation}
\mathrm{epi}(t)=
\mathbb{I}\!\Bigg[
\sum_{k=1}^{9}\mathrm{present}_k(t)\ge 5 
\ \land\ 
\exists c\in\{1,2\}:\mathrm{present}_c(t)=1
\Bigg].
\label{eq:episode-compact}
\end{equation}

This gating logic operationalizes the DSM-5 “nearly every day” rule through a moving window over daily network-derived likelihoods, analogous to sliding-window detection in traffic analytics. When at least five DSM-5 criteria, including one core criterion, are active, the router reports an \emph{episode-likely} state. 
All parameters (\(w\), \(\theta\), \(M\), \(N\)) are explicit and tunable, providing the same transparency and reproducibility expected in network-telemetry pipelines. 

\begin{figure}[H]
\centering
\resizebox{!}{.45\columnwidth}{%
    \begin{tikzpicture}[
      box/.style={draw, minimum width=7mm, minimum height=7mm},
      good/.style={fill=green!28},
      bad/.style={fill=red!22},
      lab/.style={font=\scriptsize},
      arrow/.style={-Latex, thick}
    ]
    
    % ---- Row of M example days (here we draw 14 to illustrate a generic M) ----
    \node[lab, align=center] at (-1.8,0.9) {Last $M$ days};
    \foreach \i/\c in {0/good,1/bad,2/good,3/good,4/bad,5/good,6/good,7/bad,8/good,9/good,10/good,11/bad,12/good,13/good}{
      \node[box,\c] (d\i) at (-1.4+0.8*\i,0) {};
    }
    
    % First and last day labels (parameter-consistent)
    \node[lab, above=2pt of d0] {$d=t\!-\!M\!+\!1$};
    \node[lab, above=2pt of d13] {$d=t$};
    
    % Brace under the whole window to show M
    \draw [decorate,decoration={brace,amplitude=5pt,mirror}] 
      ($(d0.south west)+(0,-0.18)$) -- ($(d13.south east)+(0,-0.18)$)
      node[midway, yshift=-10pt, lab] {$M$ consecutive days};
    
    % Legend: what a square means (uses same notation as formulas)
    \node[box,good] (gleg) at (0.2,-2.0) {};
    \node[lab, right=4pt of gleg] {$L_k(d)\ \ge\ \theta\ \Rightarrow\ I_k(d)=1$ (positive day = indication toward a DSM criterion)};
    \node[box,bad]  (bleg) at (0.2,-2.8) {};
    \node[lab, right=4pt of bleg] {$L_k(d)\ <\ \theta\ \Rightarrow\ I_k(d)=0$ (negative day = no indication toward a DSM criterion)};
    
    % Sum of positives and comparison to N
    \draw[arrow] (d13.east) -- ++(1.1,0);
    \node[box, align=right] at (3,-4.4)
      {$\displaystyle\sum_{d=t-M+1}^{t} I_k(d)\ \ge\ N\ \Rightarrow\ \mathrm{present}_k(t)=1$\\[4pt]
       otherwise $\ \mathrm{present}_k(t)=0$};
    
    \end{tikzpicture}
}
\caption[DSM-5 gate day-level visualization.]{Each square represents \emph{one day} in the last $M$ days. A day is green if the daily likelihood for criterion $k$ exceeds the threshold ($L_k(d)\ge\theta$), otherwise red. Count the green (positive) days and if at least $N$ of the last $M$ days are positive, the criterion is marked present.}
\label{fig:gate}
\end{figure}
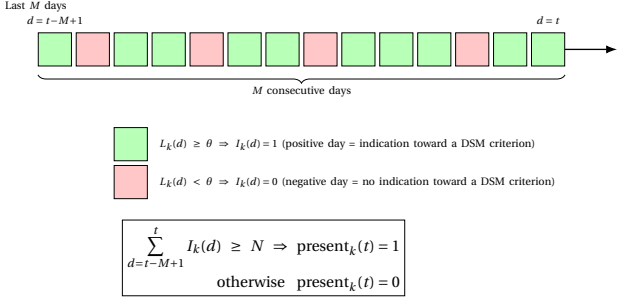

The resulting decision process can be visualized as a sliding window over daily likelihoods, where each day is marked positive or negative depending on whether its criterion likelihood exceeds the threshold~$\theta$. 
Figure~\ref{fig:gate} illustrates this temporal gating logic and how it maps continuous router-derived signals to discrete DSM-5 criterion presence.

Days with insufficient evidence are excluded from the window sum (data-quality rule), ensuring that decisions are not driven by missing or unreliable measurements. In practice, this design allows a household gateway to behave like an interpretable measurement node, aggregating packet header-level observations, applying fuzzy inference locally, and exposing clear, verifiable indicators rather than opaque scores (such as using an LLM to link features to BOMs, and BOMs to symptoms).
%=========================================================
\section{Experimental Evaluation}\label{sec:evaluation}
%=========================================================
This section evaluates whether the router‑centric prototype can (i) extract stable traffic header‑level features from raw PCAP, (ii) transform them into day‑level metrics via fixed windowing and transparent membership functions, and (iii) aggregate them into DSM‑aligned indicators using a persistence gate.

\subsection{Setup and datasets}

\noindent \textbf{Hardware.} For implementation and testing, \textit{CareNet} was evaluated on a Lenovo x64 (AMD Ryzen~7 7730U), containerized with Docker~\cite{docker} for portability, and deployed to Streamlit Community Cloud (\url{https://unisg-nef.streamlit.app/}). The platform provides approximately 0.078–2\,vCPUs, 0.69–2.7\,GB RAM, and up to 50\,GB storage per app~\cite{streamlit_manage_your_app_docs}, which suffices for the ETL, aggregation, and visualization workflows. All evaluations in this chapter are reproducible in this controlled environment, \textit{CareNet}'s \textbf{source-code is also available} \cite{nef_app_pcap_to_dsm5_public}.

\noindent \textbf{Datasets.} Two complementary packet-capture datasets were utilized, one from Jiang \etal \cite{jiang2025real}, which combines operator-side (BRAS) and customer-side (ONU) traces, achieving over 95\% application-identification accuracy. This dataset serves as a semantic link between packet-level descriptors and user activities. The second dataset, from Hjelmvik’s~\cite{hjelmvik2015hands}, offers a continuous 40-day capture with realistic, unlabeled household-like traffic. Together they cover content-aware and time-continuous dimensions needed to test \textit{CareNet’s} router-centric mapping of network metrics to behavioral indicators. %Using established public datasets ensured unbiased, ethically collected, and sufficiently rich traces resembling household environments, enabling a reproducible analysis.

\subsection{Evaluation Methodology}

Traffic is partitioned into fixed windows (\(\Delta{=}5\) min). From each window we compute counts (packets/bytes), transport shares (TCP/UDP/Other), session gaps, and DNS statistics at registrable domain (SLD/eTLD+1) level. Day‑level metrics \(x_i(d)\) are mapped to memberships \(\mu_i(d)\in[0,1]\) using triangular functions \(\mu_i(x;\ell_i,m_i,h_i)\) with optional inversion when “less = more evidence”. Per‑criterion daily likelihoods are then:
\[
L_k(d)=\sum_{i\in\mathcal{F}_k} w_i\,\mu_i(d),\quad \sum_i w_i=1,
\]
and a persistence gate \((M{=}14,\,N{=}6,\,\theta{=}0.6)\) marks criterion \(k\) “present” iff at least \(N\) of the last \(M\) days satisfy \(L_k(d)\ge\theta\). We evaluate four criteria in CareNet (C1, C2, C4, and C8) but, due to space, we focus here on C4 (sleep timing/duration) and C8 (attention/indecisiveness).

\subsection{Criteria C4 and C8 — Sleep and Attention Indicators}

\noindent \textbf{Configuration.} Criteria~C4 and~C8 represent two complementary, router-visible behaviors: diurnal rhythm (sleep timing and duration) and attention stability (concentration and decisiveness). Table~\ref{tab:c4c8-merged} lists the configured metrics, weights, and triangular membership limits. 

\begin{table}[h]
  \centering
  \caption{Configuration of parameters for C4 (Sleep timing/duration) and C8 (Difficulty concentrating/indecisiveness).}
  \label{tab:c4c8-merged}
  \scriptsize
  \setlength{\tabcolsep}{3pt}
  \renewcommand{\arraystretch}{1.05}
  \begin{tabular}{@{}p{0.46\columnwidth}ccccc@{}}
    \toprule
    \textbf{C4/C8 Feature} & $w$ & lo & mid & hi & Dir. \\
    \midrule
    \multicolumn{6}{@{}l@{}}{\textit{C4 — Sleep timing / duration}}\\
    C4\_F2 WakeAfter0400Min & 0.65 & 120 & 1085 & 1085 & $\uparrow$ \\
    C4\_F4 SleepDurationZAbs30d & 0.20 & 0.25 & 0.80 & 0.80 & $\uparrow$ \\
    C4\_F7 DaytimeIdleRatio0818 & 0.05 & 0.00 & 0.08 & 0.16 & $\uparrow$ \\
    C4\_F8 NightDayTrafficRatioBytes & 0.15 & 0.20 & 1.00 & 1.00 & $\uparrow$ \\
    \addlinespace[2pt]
    \multicolumn{6}{@{}l@{}}{\textit{C8 — Difficulty concentrating / indecisiveness}}\\
    C8\_F2 DNSBurstRatePerHour & 0.40 & 25 & 61 & 61 & $\uparrow$ \\
    C8\_F4 RepeatedQueryRatio60m & 0.40 & 0.80 & 1.00 & 1.00 & $\uparrow$ \\
    C8\_F8 MedianIKSsec & 0.20 & 0.12 & 0.22 & 0.22 & $\uparrow$ \\
    \midrule
    \multicolumn{6}{@{}l@{}}{\footnotesize MF: triangular; Dataset~B [ALL\_OTHER]; $M{=}14$, $N{=}6$, $\theta{=}0.6$, $\tau{=}1$.}\\
    \bottomrule
  \end{tabular}
  \vspace{2pt}

  \raggedright\footnotesize
  \textit{Notes:} $w$ = weight; MF = membership function; lo/mid/hi = membership limits; Dir.\ = direction ($\uparrow$ = increase indicates evidence).\\
  %.\\[-2pt]
\end{table}

\begin{figure*}[ht]
  \centering
  \begin{minipage}[t]{0.48\textwidth}
    \centering
    \includegraphics[width=\linewidth]{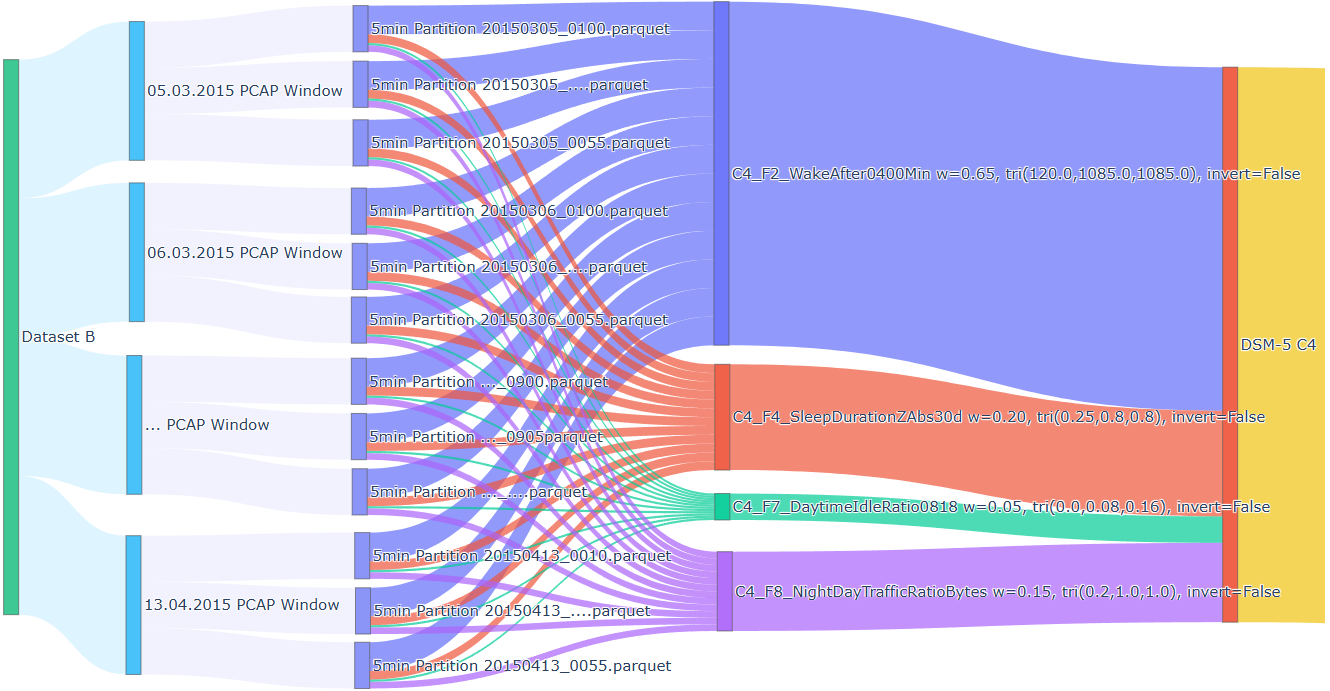}
    \caption*{\textbf{(a)} C4 — Sleep timing and duration.}
  \end{minipage}\hfill
  \begin{minipage}[t]{0.48\textwidth}
    \centering
    \includegraphics[width=\linewidth]{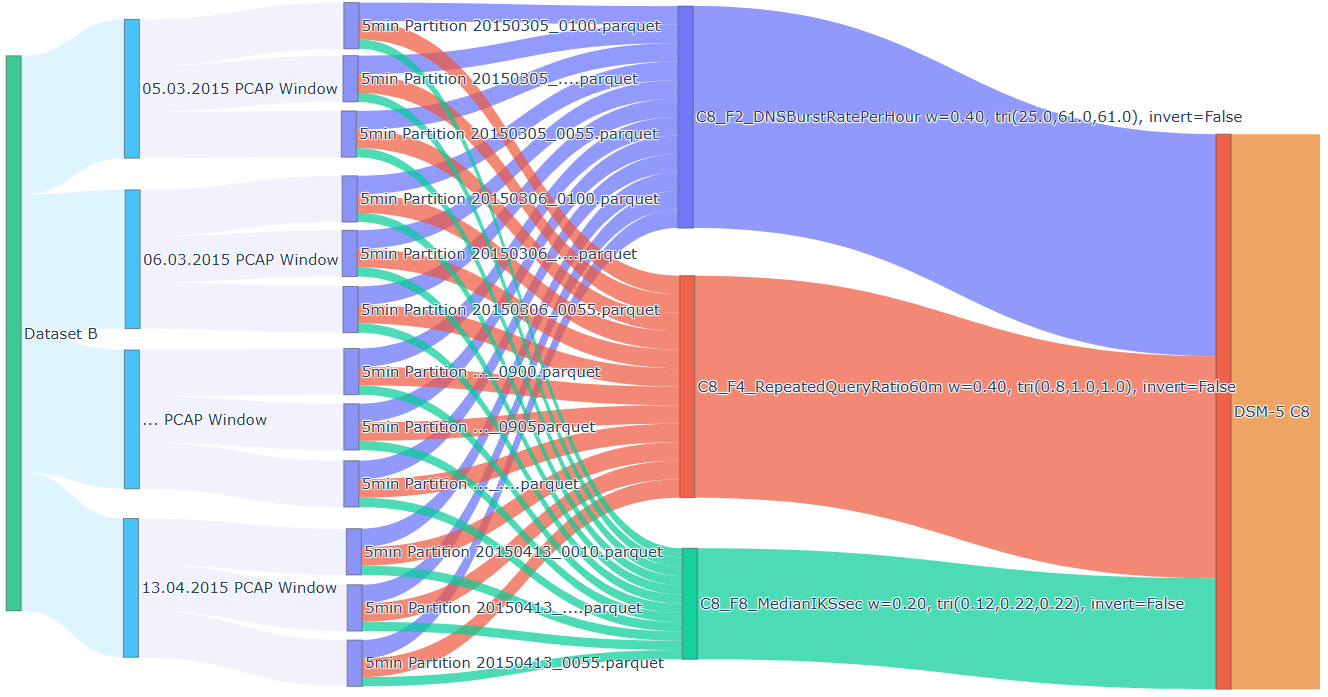}
    \caption*{\textbf{(b)} C8 — Difficulty concentrating / indecisiveness.}
  \end{minipage}
  \caption{Information flow for Criteria~C4 and~C8 showing feature extraction and aggregation into DSM-5–aligned indicators.}
  \label{fig:c4c8-sankey}
\end{figure*}

\textbf{Influence of features.} Figure~\ref{fig:c4c8-sankey} illustrates how raw packet telemetry is translated into interpretable behavioral likelihoods through the FASL pipeline. The diagrams highlight that each feature operates as a deterministic mapping from measurable network primitives, \eg packet counts, inter-arrival gaps, or DNS lookups—into bounded evidence values that together characterize either diurnal rhythm (C4) or attention stability (C8).

For C4, the dominant signal is \emph{WakeAfter0400Min} (weight 0.65), derived from the last observed network activity before 04{:}00 local time. It anchors the detection of delayed or fragmented sleep patterns by identifying nights where activity persists abnormally late. The secondary features refine this estimate: \emph{SleepDurationZAbs30d} quantifies how much total nightly inactivity deviates from the 30-day baseline, \emph{DaytimeIdleRatio0818} measures the proportion of idle time during the main activity window (08{:}00–18{:}00), and \emph{NightDayTrafficRatioBytes} compares nocturnal and daytime byte volumes. Together they approximate a digital circadian profile purely from transport- and session-level metadata, showing that even coarse temporal packet aggregates can encode sleep-wake regularity without inspecting payloads.

For C8, the decisive features are \emph{DNSBurstRatePerHour} and \emph{RepeatedQueryRatio60m} (weights 0.40 each). The first measures how often multiple distinct second-level domains are contacted in rapid succession, indicating short attention spans or topic switching; the second counts repeated DNS queries for identical domains within an hour, which may reflect checking or uncertainty behaviors. Both are observable directly from resolver traffic and require no flow reconstruction. The auxiliary feature \emph{MedianIKSsec} (weight 0.20) estimates inter-keystroke gaps inferred from packet bursts in interactive sessions, providing an orthogonal, human-interaction dimension. %Their joint behavior captures instability in online focus using only timing and naming metadata—no content or user identifiers.

\begin{figure}[t]
  \centering
  \begin{minipage}{\columnwidth}
    \centering
    \includegraphics[width=\linewidth]{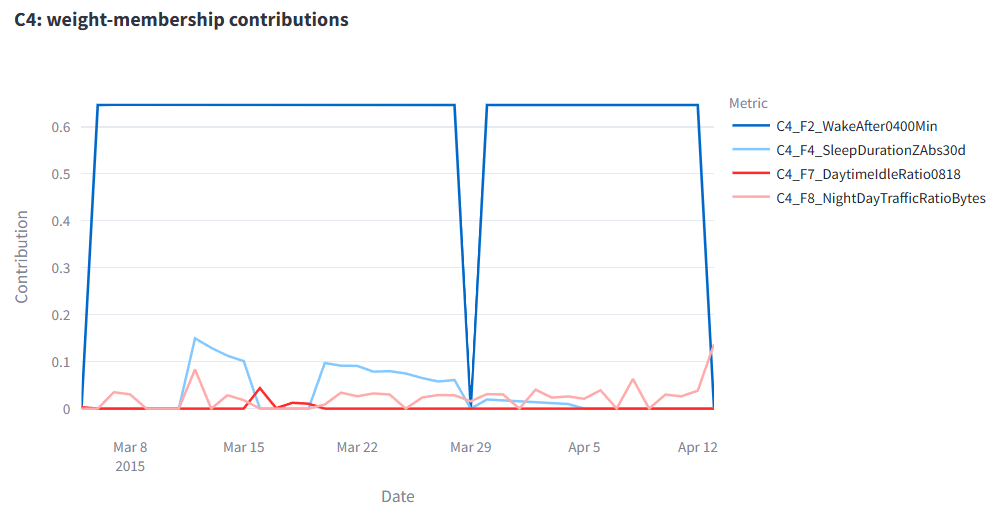}
    \caption*{\textbf{(a)} C4 — weight–membership over time.}
  \end{minipage}
  \vspace{2pt}
  \begin{minipage}{\columnwidth}
    \centering
    \includegraphics[width=\linewidth]{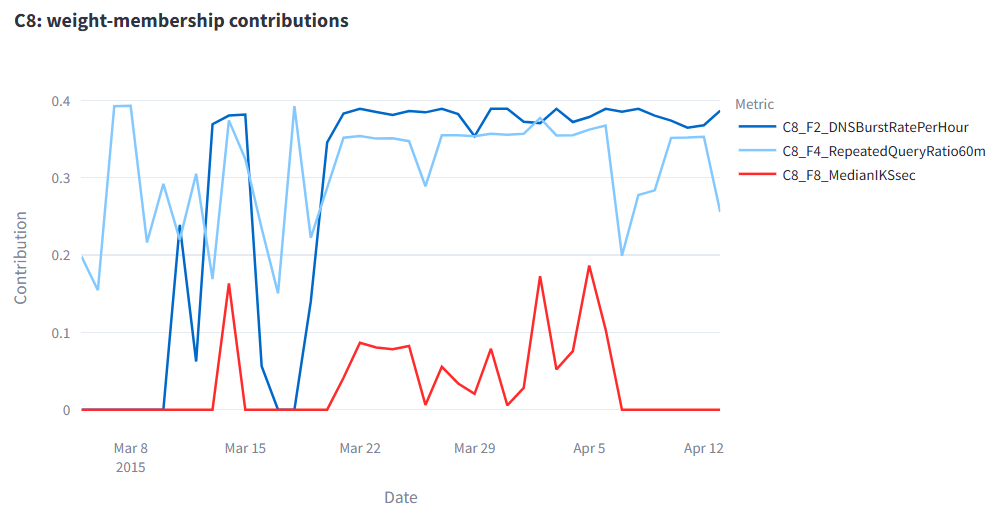}
    \caption*{\textbf{(b)} C8 — weight–membership over time.}
  \end{minipage}
  \caption{Weight–membership contributions for C4 and C8.}
  \label{fig:c4c8-contrib}
\end{figure}

\noindent \textbf{Membership functions.} The contribution stacks in Figure~\ref{fig:c4c8-contrib} visualize how each feature’s membership value contributes to the overall daily likelihoods. These plots make the internal mechanics of the FASL fusion explicit: all metrics are first mapped through bounded triangular functions, then linearly combined according to their configured weights (\cf Table \ref{tab:c4c8-merged}). These resulting traces show how much evidence each metric provides on each day, making the additive nature of the inference directly observable.

For C4, the dominant feature \emph{WakeAfter0400Min} drives most of the accumulated membership, reflecting persistent late-night activity visible at the packet level. The other features—\emph{SleepDurationZAbs30d}, \emph{DaytimeIdleRatio}, and \emph{Night/DayRatio}—contribute short pulses that refine this baseline by accounting for sleep duration irregularities, daytime inactivity, and nocturnal traffic surges. Together they model both persistent and transient disruptions in digital circadian rhythm. The visual plateau in Figure~\ref{fig:c4c8-contrib}(a) demonstrates that the membership saturates near its upper bound for extended periods, a behavior desirable in fuzzy logic because it prevents single anomalies from dominating the criterion.

For C8, the decomposition in Figure~\ref{fig:c4c8-contrib}(b) shows how attention-related features interact over time. Early in the sequence, peaks in \emph{RepeatedQueryRatio60m} indicate short-lived bursts of redundant DNS lookups, followed by a sustained plateau dominated by \emph{DNSBurstRatePerHour}. This reflects periods of rapid, topic-switching network behavior, as seen in the frequent resolver activity across distinct domains. %The auxiliary metric \emph{MedianIKSsec} introduces intermittent contributions from a separate human-interaction channel; these pulses verify that attention-related evidence is not tied to one modality but corroborated across independent sources of telemetry.

\begin{figure}[t]
  \centering
  \begin{subfigure}{\columnwidth}
    \centering
    \includegraphics[width=\linewidth]{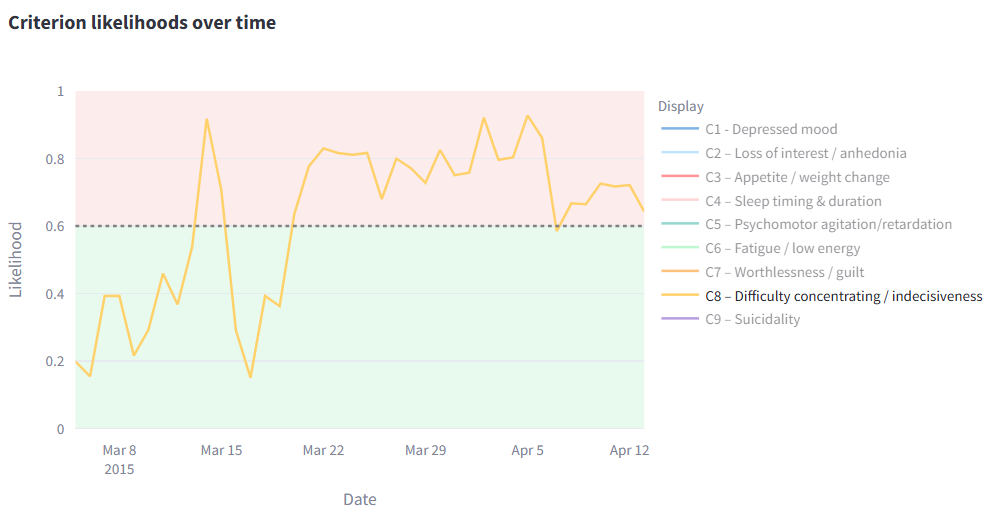}
    \caption{C8 — criterion likelihood over time. Multi-metric alignment produces sustained periods above the gate; isolated fluctuations only nudge the curve.}
    \label{fig:c8-like-sub}
  \end{subfigure}
  \vspace{2pt}
  \begin{subfigure}{\columnwidth}
    \centering
    \includegraphics[width=\linewidth]{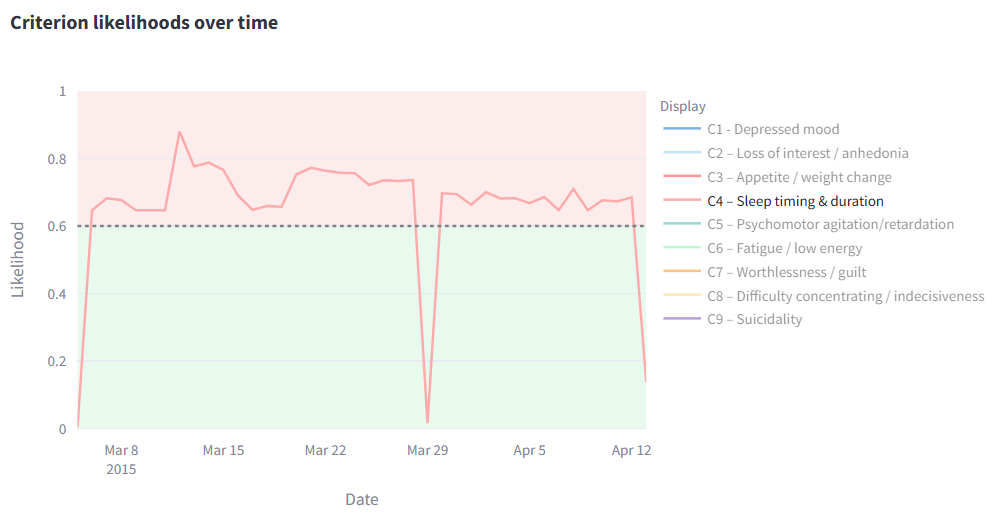}
    \caption{C4 — criterion likelihood over time. Long stretches above the gate reflect a saturated wake-after-04:00 signal with small refinements from the other metrics and a single reset day near the end of March.}
    \label{fig:c4-like-sub}
  \end{subfigure}
  \caption{Daily criterion likelihoods for C8 and C4 after FASL aggregation and DSM-style gating.}
  \label{fig:c4c8-likelihood}
\end{figure}

\noindent \textbf{Criterion likelihood.}  Figure~\ref{fig:c4c8-likelihood} presents the aggregated daily likelihoods for Criteria~C8 and~C4 after combining all weighted feature memberships through FASL. Each curve represents the day-level probability that the corresponding behavioral criterion is active, filtered by the temporal gate parameters \((M{=}14,\,N{=}6,\,\theta{=}0.6)\).

For C8, the likelihood trace shows clear multi-day periods where DNS-related metrics align above the decision threshold, indicating consistent evidence of attentional fragmentation. These plateaus arise when both \emph{DNSBurstRatePerHour} and \emph{RepeatedQueryRatio60m} remain elevated for consecutive days, while intermittent keystroke-gap pulses produce smaller transient increases. This shows the gate acts as a temporal integrator, requiring multiple evidence sources to confirm anomalies within a window.

In contrast, C4 exhibits extended intervals above the gate dominated by the persistent \emph{WakeAfter0400Min} signal, which reflects continuous late-night network activity in Dataset~B. Secondary features such as \emph{SleepDurationZAbs30d}, \emph{DaytimeIdleRatio}, and \emph{Night/DayRatio} modulate these plateaus, adding sensitivity to short-term variations in daily rhythm. The stable sections above \(\theta\) therefore represent corroborated, time-consistent deviations rather than single-day outliers.

\subsection{The User Journey}

The final overview aggregates all DSM-5 gate outcomes into a unified, interpretable interface within \emph{CareNet}. It reports three key indicators derived from the same temporal parameters \((M{=}14,\,N{=}6,\,\theta{=}0.6,\,\tau{=}1)\): (i) overall episode likelihood, (ii) core-symptom activation, and (iii) the count of criteria currently exceeding their gates within the 14-day window. This stage closes the analytical loop by translating the per-criterion likelihoods into a system-level diagnostic summary.

\begin{figure}[h]
  \centering
  \includegraphics[width=\columnwidth]{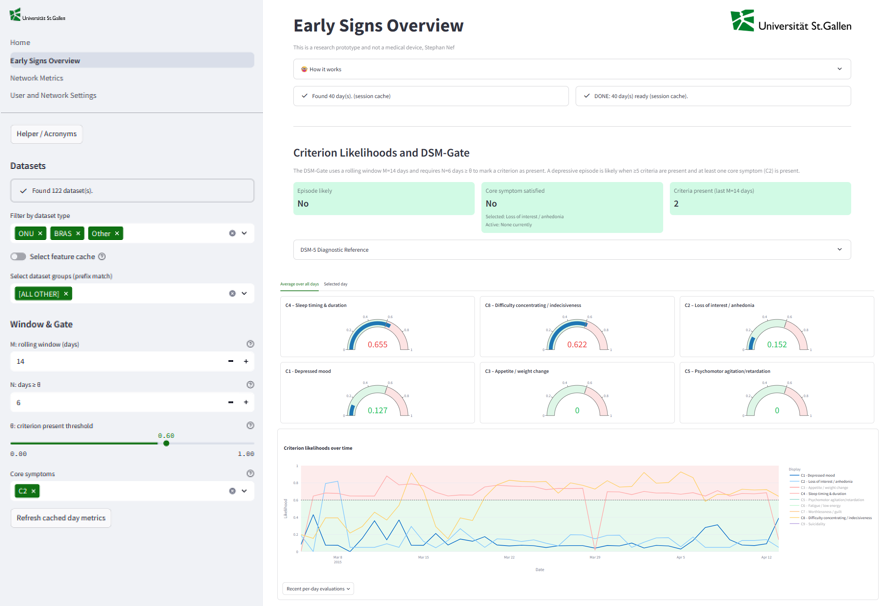}
  \caption{Entry point \emph{Early Signs Overview}. The DSM-5 gate and triage elements are summarized at a glance for quick interpretation.}
  \label{fig:entry-overview}
\end{figure}

Figure~\ref{fig:entry-overview} introduces this final layer of analysis, where the DSM-style gate is applied globally to produce an at-a-glance behavioral snapshot. In the evaluated configuration, no episode is detected and no core symptom (C1, C2) exceeds its gate, while two non-core criteria, C4 (sleep) and C8 (attention), remain persistently elevated. This outcome reproduces the per-criterion trends observed earlier: stable deviations in circadian and attentional metrics without concurrent changes in mood or motivation. The gate follows DSM-5 logic by requiring multiple corroborating symptoms within a fixed time horizon, translating clinical persistence into a computationally verifiable condition.

This rule-based formulation counts active criteria \(k\) with \(L_k(d)\geq\theta\) for at least \(N\) of the past \(M\) days and reports both the instantaneous and aggregated state. Because the configuration and thresholds are explicit, every high-level decision can be traced back to specific metrics and timestamps, ensuring auditability and reproducibility. From a systems viewpoint, this layer acts as the final aggregation of a multi-stage signal-processing chain: packet-level measurements → daily features → criterion likelihoods → interpretable episode-level indicators.

\begin{figure}[t]
  \centering
  \includegraphics[width=1\columnwidth]{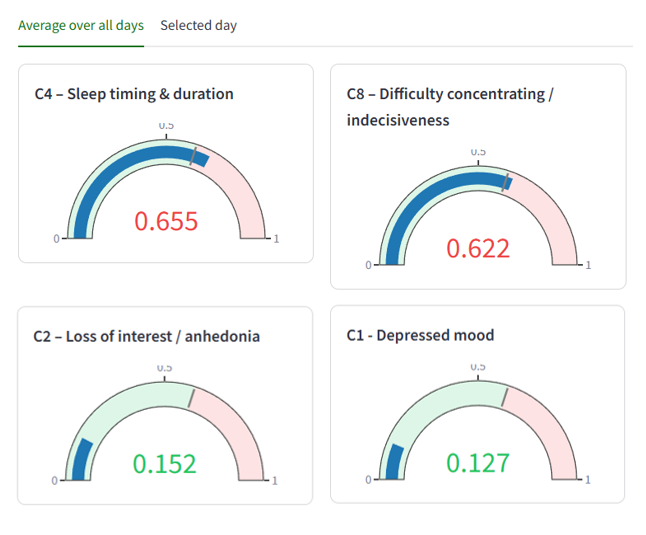}
  \caption{Sorted gauges overview. Criteria are ranked by average likelihood, highlighting persistent elevations in C4 and C8.}
  \label{fig:gauges-sorted}
\end{figure}

\begin{figure}[t]
  \centering
  \includegraphics[width=\columnwidth]{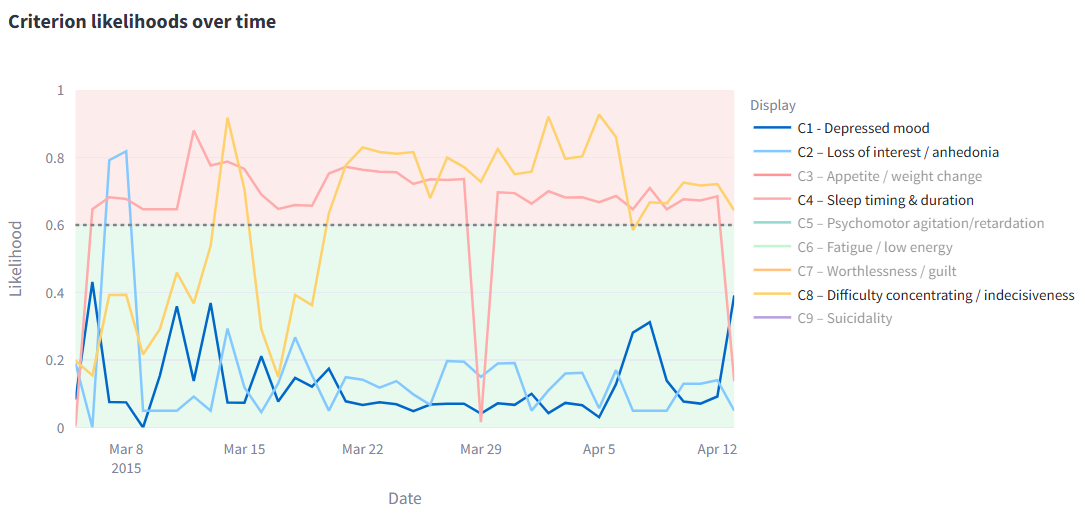}
  \caption{All criteria over time. Co-movement of C4 and C8 dominates, while C1 and C2 remain near baseline.}
  \label{fig:all-crit-time}
\end{figure}

Figures~\ref{fig:gauges-sorted} and~\ref{fig:all-crit-time} depict the second layer of interpretation. The gauge panel aggregates mean likelihoods across the observation window, ranking C4 (0.655) and C8 (0.622) highest. The multi-criterion time series confirms this relationship, showing parallel trajectories of sleep and attention features with minor, short-lived excursions for mood-related criteria. Together, these visualizations show the scalability of the approach, from per-packet evidence to daily metrics and daily metrics to criterion likelihoods, and finally to explainable depression symptoms via the dashboard.

\section{Discussion and Limitations}
\label{sec:discussion}

%With only packet headers, we computed reproducible metrics, mapped them to simple membership functions, and combined them with an auditable gate that mirrors DSM‑5 structure. 
The evaluation shows three aspects that matter in practice. First, the method is \emph{explainable by design}. Each metric comes with an explicit baseline (\texttt{lo}/\texttt{mid}/\texttt{hi}), a visible membership curve and a time-series of outputs. Users can see how moving a limit or changing a weight changes the indicator. Second, the approach is \emph{person‑centric}. Healthy baselines are set per user and outliers above \texttt{hi} are intentionally ignored, which reduces noise while keeping unusual days visible in the raw plots. Third, the DSM‑style gate aggregates day‑level likelihoods in a way that “feels familiar” to clinicians: it requires corroboration across metrics and at least one core symptom before declaring that an episode is likely.

\noindent \textbf{Deterministic, auditable, and open-source.} This is a key aspect that makes our output explainable. Every indicator is explained by a visible rule set rather than a black‑box model. All packet‑level primitives (\eg inter‑arrival structure, direction/volume, DNS activity) map one‑to‑one to daily features, and triangular memberships convert deviations into bounded evidence; and the persistence gate \((M,N,\theta,\tau)\) enforces temporal corroboration. Our introduced FASL pipeline behaves like a signal‑processing stack down-sampling high‑rate traffic into low‑frequency behavioral metrics with explicit sensitivity and thresholds. %In addition, reproducibility is ensured via a containerized app and full code release \cite{nef_app_pcap_to_dsm5_public}. %We also provide the evaluation inputs used in this paper, enabling independent verification and extension.

\noindent \textbf{Privacy considerations.} Both network metadata and mental-health indicators are extremely sensitive information. The edge-first architecture ensures that all computation and data storage stay within the household. Only packet headers and derived, day-level features are used, while payloads are never inspected and no data leave the home by default. This approach minimizes risk, aligns with data-minimization principles, and increases user trust. %The strict privacy requirements shaped the design in a positive way. By focusing on content-agnostic features such as timestamps, directions, and hostnames, every step of the analysis became interpretable and straightforward to verify.

\noindent \textbf{Limitations: What a router can and cannot observe.} A network‑centric view can provide several meaningful behavioral metrics: diurnal rhythm (\eg night/day balance, quiet‑hour structure), but not all DSM-5 depression indicators are reliably inferable from network traffic alone (\eg mood valence, guilt/worthlessness, ideation content, detailed psychomotor signs). In addition, while current likelihoods and weights are currently calibrated against statistical baselines, note that these require clinical/behavioral ground-truth (\eg PHQ‑9, actigraphy \cite{kroenke2001phq}). In this regard, \textit{CareNet} allows to tune each weight/parameter in a way that it could be extended to complement existing clinical practice (rather than replacing clinical assessment).

\section{Summary and Future Work}
\label{sec:final_considerations}

This paper presented \emph{CareNet}, a router-centric framework that transforms household network telemetry into interpretable behavioral indicators aligned with DSM-5 symptom domains. \emph{CareNet} presents a complete edge-based pipeline, from packet capture and feature extraction to fuzzy aggregation and DSM-style gating, showing that clinically relevant temporal patterns can be inferred from header-level data alone. 

Future work will focus on validating these indicators against clinical ground truth, extending the framework to additional DSM-5 domains, and automating parameter tuning to adapt thresholds and weights dynamically across heterogeneous household environments.

\section{Ethical Statement}
\label{sec:ethical}

\textit{CareNet} is based on Stephan's master thesis at the University of St. Gallen. AI tools were used for spelling, grammar, style, and code assistance in prototype development. The conceptualization, methodological design, and interpretation of results originated from the authors, while AI tools served only as auxiliary instruments for language refinement and implementation support.

%====================================================

%\bibliographystyle{IEEEtran}
% <OR> manually copy in the resultant .bbl file set second argument of \begin to the number of references (used to reserve space for the reference number labels box)

%====================================================


%\section*{Acknowledgments} 
%This work has been supported by 

%\IEEEtriggeratref{16}
% The "triggered" command can be changed if desired: \IEEEtriggercmd{\enlargethispage{-5in}}

%balance
%\bibliographystyle{IEEEtran}
\bibliographystyle{IEEETranS}
%\balance
\bibliography{bib/main.bib}

%\clearpage
% Start of appendix
\appendix

\section*{Repository Access and Reproducibility}

\begin{figure}[H]
  \centering
  \includegraphics[width=\columnwidth]{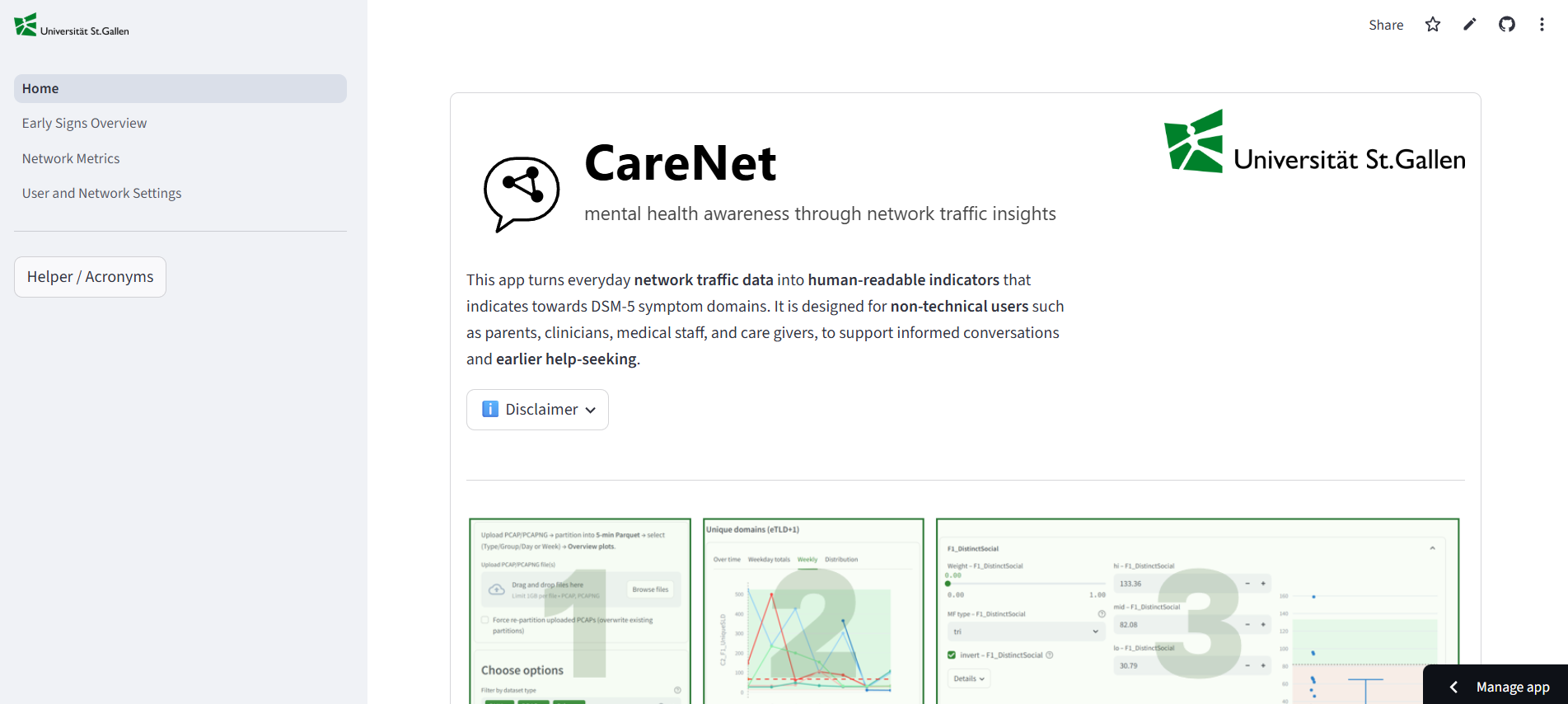}
  \caption[Landing page.]{CareNet's landing page.}
  \label{fig:journey-home}
\end{figure}

\textbf{Source code and live deployment.} All components required to reproduce the results presented are publicly available. The following notes describe where to find the source code, how to run the application locally, and how the repository is structured, and build the system in Figure \ref{fig:journey-home}.

\medskip
\noindent
 
The full implementation of \textit{CareNet} is hosted on GitHub at  
\href{https://github.com/nefste/app_pcap_to_dsm5_public/tree/main/app}{\texttt{nefste/app\_pcap\_to\_dsm5\_public}}.  
It includes the complete Streamlit dashboard, metric definitions, ETL utilities, and configuration files used in the evaluation.  
A live instance is available at \url{https://unisg-nef.streamlit.app/}. Access credentials can be obtained directly from the author upon request.

\medskip
\noindent \textbf{Running locally.} In addition, \textit{CareNet} can be executed either through Docker or within a Python virtual environment. The following commands reproduce the full setup:

\begin{lstlisting}[language=bash,basicstyle=\ttfamily\small, caption={Run locally using Docker}]
cd app
docker compose up --build
# Application available at http://localhost:8501
\end{lstlisting}

\begin{lstlisting}[language=bash,basicstyle=\ttfamily\small, caption={Run locally using a Python virtual environment}]
cd app
python -m venv ..\.venv
..\.venv\Scripts\activate   # on Windows (adjust for macOS/Linux)
pip install -r requirements.txt
streamlit run 00_Home.py
\end{lstlisting}

\noindent
If additional utilities are required, install repository-level dependencies using  
\texttt{pip install -r ../requirements.txt}.  
When \texttt{libpcap} is unavailable, set \texttt{SCAPY\_USE\_LIBPCAP=no}; the application reads PCAP files from disk and does not require live packet capture.

\medskip
\noindent
\textbf{Repository structure.}  
The main files and directories of the application are summarized below. Each entry lists the purpose of the corresponding component and its role in the overall system.

\begin{description}\footnotesize
  \item[A1] \texttt{app/00\_Home.py} — Landing page with disclaimer and general overview of the dashboard.
  \item[A2] \texttt{app/pages/01\_Early\_Signs\_Overview.py} — Loads and normalizes Parquet data; computes and visualizes FASL and DSM-Gate outputs.
  \item[A3] \texttt{app/pages/02\_Network\_Metrics.py} — Defines per-criterion metric computations and corresponding UI views.
  \item[A4] \texttt{app/pages/03\_User\_and\_Network\_Settings.py} — Manages profiles, IP mappings, uploads, and ETL partitioning.
  \item[A5] \texttt{app/metrics/common.py} — Contains helper functions for sessionization (\texttt{sessions\_from\_timestamps}), domain parsing (eTLD+1), and data formatting.
  \item[A6] \texttt{app/metrics/base\_features.py} — Generates daily base features such as \textit{LateNightShare}, \textit{LongestInactivityHours}, and IS/IV ratios.
  \item[A7] \texttt{app/metrics/criterion1.py--criterion9.py} — Implements DSM-5 criterion-specific metrics and their respective likelihood computations.
  \item[A8] \texttt{app/auto\_tune.py} — Optional module for auto-tuning triangular memberships and feature weights.
  \item[A9] \texttt{app/processed\_parquet/<dataset>/} — Stores five-minute Parquet files linking ETL and analytics stages (\texttt{<dataset>\_\_YYYYMMDD\_HHMM.parquet}).
\end{description}

\noindent
%Together, these components form a complete and reproducible pipeline for router-centric sensing, from packet parsing and feature extraction to behavioral aggregation and visualization.

\section*{Loading Traces and Configuring User Profiles}

\textit{CareNet} supports interactive exploration of packet-capture datasets directly within the dashboard. 

\begin{figure}[H]
  \centering
  \includegraphics[width=\columnwidth]{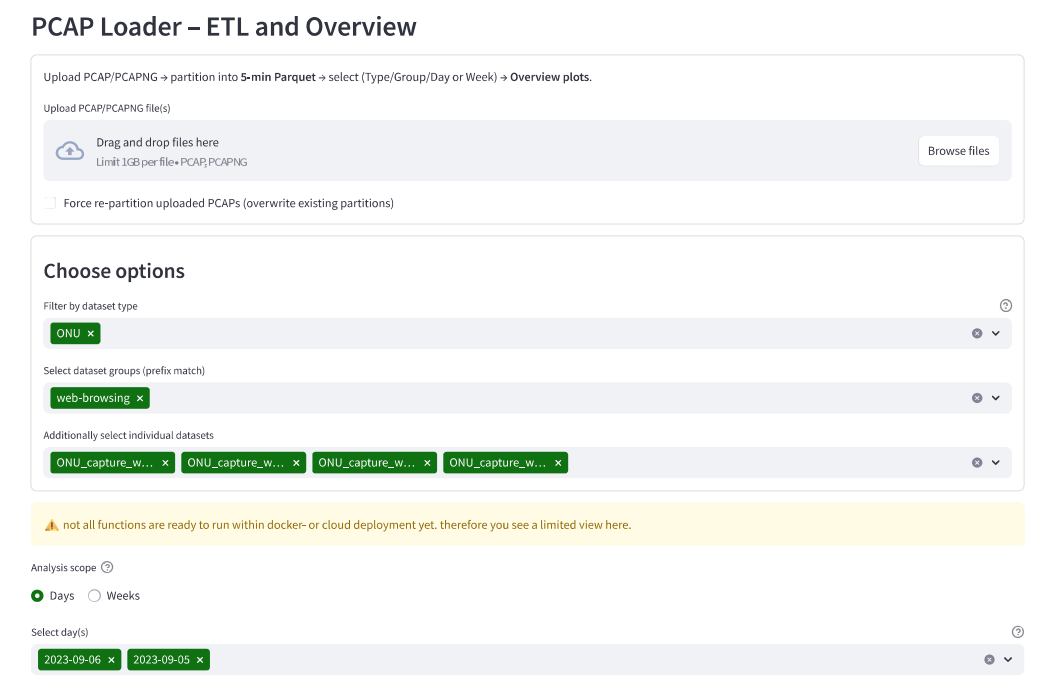}
  \caption[App configuration interface.]{Configuration page for selecting datasets and analysis periods in the \textit{CareNet} dashboard.}
  \label{fig:user-config}
\end{figure}

After launching the application, users can open the \textit{User and Network Settings} page to configure which PCAPs to load and how they are partitioned (\cf Figure \ref{fig:user-config}). 
The interface allows selecting one or more dataset groups such as \texttt{BRAS}, \texttt{ONU}, or \texttt{snort.log}, as well as thematic subsets (e.g., \emph{video}, \emph{web-browsing}, or \emph{gaming}). 

%A \emph{Raw Data Preview} pane displays basic packet and domain statistics before processing, and the corresponding pre-aggregated CSV file can be downloaded for inspection or offline analysis. 

\begin{figure}[h]
  \centering
  \includegraphics[width=0.70\columnwidth]{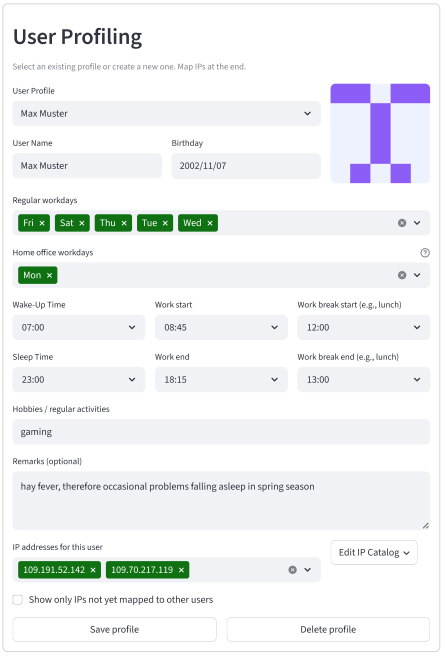}
  \caption[User profiling and IP mapping.]{User profiling and IP mapping used by direction-aware metrics and interpretations.}
  \label{fig:journey-profile}
\end{figure}

It is also possible to maintain profiles (wake/sleep habits, workdays, notes) and map local IPs to users (Figure~\ref{fig:journey-profile}). This mapping activates lightweight direction heuristics (private$\to$public $\approx$ upstream, public$\to$private $\approx$ downstream), improving “active vs.~passive” features without content access. %Profiles have not yet been used within the FASL calculation (Figure~\ref{fig:journey-profile}).

\end{document}